\documentclass[12pt]{article}
\usepackage{amsmath,amsfonts,amssymb}
\usepackage[dvips]{graphicx}
\usepackage{color,empheq}

\def\red#1{{\color{red} #1}}
\def\affil#1{\\ {\normalsize #1}}  
\begin{document}

\def\prg#1{\medskip\noindent{\bf #1}}  \def\ra{\rightarrow}
\def\lra{\leftrightarrow}              \def\Ra{\Rightarrow}
\def\nin{\noindent}                    \def\pd{\partial}
\def\dis{\displaystyle}
\def\grl{{GR$_\Lambda$}}               \def\Lra{{\Leftrightarrow}}
\def\cs{{\scriptstyle\rm CS}}          \def\ads3{{\rm AdS$_3$}}
\def\Leff{\hbox{$\mit\L_{\hspace{.6pt}\rm eff}\,$}}
\def\bull{\raise.25ex\hbox{\vrule height.8ex width.8ex}}
\def\ric{{Ric}}                        \def\tric{{(\widetilde{Ric})}}
\def\tmgl{\hbox{TMG$_\Lambda$}}        \def\phb{\phantom{\Big|}}
\def\Lie{{\cal L}\hspace{-.7em}\raise.25ex\hbox{--}\hspace{.2em}}
\def\sS{\hspace{2pt}S\hspace{-0.83em}\diagup}   \def\hd{{^\star}}
\def\dis{\displaystyle}                \def\ul#1{\underline{#1}}
\def\mb#1{\hbox{{\boldmath $#1$}}}     \def\grp{{GR$_\parallel$}}
\def\irr#1{^{(#1)}}                    \def\ric{{Ric}}

\def\hook{\hbox{\vrule height0pt width4pt depth0.3pt
\vrule height7pt width0.3pt depth0.3pt
\vrule height0pt width2pt depth0pt}\hspace{0.8pt}}
\def\semidirect{\;{\rlap{$\supset$}\times}\;}
\def\bm#1{\hbox{{\boldmath $#1$}}}
\def\ir#1{{}^{(#1)}}  \def\inn{\hook}

\def\G{\Gamma}         \def\S{\Sigma}        \def\L{{\mit\Lambda}}
\def\D{\Delta}         \def\Th{\Theta}
\def\a{\alpha}         \def\b{\beta}         \def\g{\gamma}
\def\d{\delta}         \def\m{\mu}           \def\n{\nu}
\def\th{\theta}        \def\k{\kappa}        \def\l{\lambda}
\def\vphi{\varphi}     \def\ve{\varepsilon}  \def\p{\pi}
\def\r{\rho}           \def\Om{\Omega}       \def\om{\omega}
\def\s{\sigma}         \def\t{\tau}          \def\eps{\epsilon}
\def\nab{\nabla}       \def\btz{{\rm BTZ}}   \def\heps{\hat\eps}
\def\bt{{\bar t}}      \def\br{{\bar r}}     \def\bth{{\bar\theta}}
\def\bvphi{{\bar\vphi}}  \def\bO{{\bar O}}   \def\bx{{\bar x}}
\def\by{{\bar y}}      \def\bom{{\bar\om}}
\def\aa{{\bar a}}     \def\bb{{\bar b}}
\def\tphi{{\tilde\vphi}} \def\tt{{\tilde t}}

\def\tG{{\tilde G}}   \def\cF{{\cal F}}      \def\bH{{\bar H}}
\def\cL{{\cal L}}     \def\cM{{\cal M }}     \def\cE{{\cal E}}
\def\cH{{\cal H}}     \def\hcH{\hat{\cH}}    \def\hA{\hat{A}}
\def\cK{{\cal K}}     \def\hcK{\hat{\cK}}    \def\cT{{\cal T}}
\def\cO{{\cal O}}     \def\hcO{\hat{\cal O}} \def\cV{{\cal V}}
\def\cE{{\cal E}}     \def\cR{{\cal R}}      \def\hR{{\hat R}{}}     \def\hL{{\hat\L}}     \def\tom{{\tilde\omega}}
\def\tb{{\tilde b}}   \def\tA{{\tilde A}}    \def\tv{{\tilde v}}
\def\tT{{\tilde T}}   \def\tR{{\tilde R}}    \def\tcL{{\tilde\cL}}
\def\hy{{\hat y}\hspace{1pt}}  \def\tcO{{\tilde\cO}}
\def\bA{{\bar A}}     \def\bB{{\bar B}}      \def\bC{{\bar C}}
\def\bG{{\bar G}}     \def\bD{{\bar D}}      \def\bH{{\bar H}}
\def\bK{{\bar K}}     \def\bL{{\bar L}}

\def\rdc#1{\hfill\hbox{{\small\texttt{reduce: #1}}}}
\def\chm{\checkmark}  \def\chmr{\red{\chm}}
\def\pgp{\text{PG$^{+}$}}             \def\pgm{\text{PG$^{-}$}}
\def\kadsp{\text{Kerr-AdS{}$^{+}$}}   \def\kads{\text{Kerr-AdS}}
\def\kadsm{\text{Kerr-AdS$^{-}$}}     \def\bS{{\bar S}}
\def\gr{\textsc{GR}}

\def\nn{\nonumber}                    \def\vsm{\vspace{-9pt}}
\def\be{\begin{equation}}             \def\ee{\end{equation}}
\def\ba#1{\begin{array}{#1}}          \def\ea{\end{array}}
\def\bea{\begin{eqnarray} }           \def\eea{\end{eqnarray} }
\def\beann{\begin{eqnarray*} }        \def\eeann{\end{eqnarray*} }
\def\beal{\begin{eqalign}}            \def\eeal{\end{eqalign}}
\def\lab#1{\label{eq:#1}}             \def\eq#1{(\ref{eq:#1})}
\def\bsubeq{\begin{subequations}}     \def\esubeq{\end{subequations}}
\def\bitem{\begin{itemize}}           \def\eitem{\end{itemize}}
\renewcommand{\theequation}{\thesection.\arabic{equation}}

\title{Thermodynamics of Riemannian Kerr-AdS black holes\\
       in Poincar\'e gauge theory}

\author{M. Blagojevi\'c and B. Cvetkovi\'c\footnote{
        Email addresses: \texttt{mb@ipb.ac.rs, cbranislav@ipb.ac.rs}}
\affil{Institute of Physics, University of Belgrade,
                      Pregrevica 118, 11080 Belgrade, Serbia} }

\date{}
\maketitle

\begin{abstract}
A Hamiltonian approach to black hole entropy is used to study Riemannian Kerr-AdS solutions in the general, parity-violating  Poincar\'e gauge theory. Entropy and the asymptotic charges are entirely determined by the parity-even sector of the theory, whereas the parity-odd contributions vanish. Entropy is found to be proportional to the horizon area, and the first law of black hole thermodynamics is confirmed.
\end{abstract}

\section{Introduction}
\setcounter{equation}{0}

From the experience with general relativity (GR), we know that exact solutions play an essential role in the physical interpretation of gravitational theories \cite{x1}. In  Poincar\'e gauge theory (PG), where both the torsion $T^i$ and the curvature $R^{ij}$ define the gravitational dynamics \cite{x2}, exact solutions have been  often constructed by suitably ``incorporating" torsion into the known solutions of GR. In particular, an advanced technique was used by Baekler et al. \cite{x3} to construct a Kerr-AdS black hole with propagating torsion, in the sector of \emph{parity-invariant} PG Lagrangians. Recently, Obukhov \cite{x4} made one more step in the same direction by extending the construction to the most general, \emph{parity-violating} PG \cite{x5}. Thus, there are at least three versions of Kerr-AdS spacetimes, one in GR and the other two in PG; they possess the same metric but live in different dynamical setups.

Investigations of black hole thermodynamics have given rise to a deeper insight into both the classical and quantum nature of gravity \cite{x6}. In the 1990s, classical black hole entropy was most compactly described as the diffeomorphism  \emph{Noether charge on horizon} \cite{x7,x8,x9}. In a recently proposed Hamiltonian approach to black hole entropy \cite{x10}, the same idea was extended to PG, where diffeomorphisms are replaced by the Poincar\'e gauge symmetry. It offers a unified description of the asymptotic charges (energy and angular momentum) and entropy of black holes \emph{with or without torsion}, in terms of certain boundary integrals at infinity and on horizon, respectively. The approach was successfully applied to Kerr-AdS thermodynamics in GR \cite{x11}, as well as to parity-invariant and general PG models \cite{x12,x13}. In the present work, we extend the Hamiltonian analysis to the class of \emph{Riemannian Kerr-AdS solutions} in the general PG. A comparison to the results found in Riemannian theories with quadratic curvature Lagrangians \cite{x8,x14}, as well as in the general PG models \cite{x13}, reveals how black holes react to different dynamical frameworks.

The paper is organized as follows. In section \ref{sec2}, we give a short account of the Hamiltonian approach to black hole thermodynamics in the context of the general, parity-violating PG. In section \ref{sec3}, we introduce the tetrad formulation of the Kerr-AdS geometry, needed in the Hamiltonian analysis, and discuss limitations of Boyer-Lindquist coordinates. Sections \ref{sec4} and \ref{sec5} contain the main results of the paper---derivation of the asymptotic charges and entropy for Riemannian Kerr-AdS blck holes in PG. Section \ref{sec6} is devoted to discussion.

Our basic notation is the same as in Refs. \cite{x11,x13}. The latin indices $(i,j,\dots)$ refer to the local Lorentz frame, the greek indices $(\m,\n,\dots)$ refer to the coordinate frame,  $b^i$ is the orthonormal coframe (tetrad) and  $\om^{ij}$ is a  metric compatible connection (1-forms), $h_i$ is the dual basis (frame) such that $h_i\inn b^k = \d_i^k$, and $\eta_{ij}=(1,-1,-1,-1)$ is the local Lorentz metric. The wedge symbol in the exterior products is omitted, the volume 4-form is $\heps = b^0b^1b^2b^3$, the Hodge dual of a form $\a$ is denoted by $\hd\a$, with $\hd 1=\heps$, and the totally antisymmetric symbol $\ve_{ijmn}$ is normalized to $\ve_{0123}=+1$.

\section{Black hole entropy as a boundary term}\label{sec2}
\setcounter{equation}{0}

In this section, we give a short account of the Hamiltonian approach to the entropy of black holes, restricted to the class of Riemannian solutions in the general PG, see Refs. \cite{x11,x13}.

We begin by recalling some geometric aspects of PG.
The gravitational field is described by two independent dynamical variables, the tetrad $b^i$ and the metric compatible spin connection $\om^{ij}$ (1-forms), which are associated to the translation and the Lorentz subgroups of the Poincar\'e group, respectively. The corresponding field strengths are the torsion $T^i=d b^i+\om^i{}_k b^k$ and the curvature $R^{ij}:=d\om^{ij}+\om^i{}_k\om^{kj}$ (2-forms), and the underlying spacetime structure is characterized by a Riemann--Cartan geometry.

The general PG dynamics is determined by a Lagrangian $L_G(b^i,T^i,R^{ij})$ which is at most quadratic in the field strengths and contains both even and odd parity modes. In this work, we are interested in the class of vacuum solutions with \emph{vanishing torsion}. Their dynamics is \emph{effectively} described by a simplified Lagrangian without torsion invariants,
\bea
L_G&=& -\hd(a_0R+\aa_0 X+2\L_0)                                        \nn\\
   && +\frac{1}{2}R^{ij}\sum_{n=1}^6\big[\hd\big(b_n\ir{n}R_{ij}\big)
                                        +\bb_n\ir{n}R_{ij}\big]\, .  \lab{2.1}
\eea
Here, $(a_0,\L_0,b_n)$ and $(\aa_0,\bb_n)$ are the Lagrangian parameters in the parity even and odd sectors, respectively, $\hd R=\hd(b_ib_j)R^{ij}$ is he Einstein-Hilbert and $\hd X=(b_ib_j)R^{ij}$ the Holst term \cite{x15}, and $\ir{n}R^{ij}$ are irreducible parts of the curvature \cite{x10}; for $T^i=0$, only those for $n=(1,4,6)$ are nonvanishing. The gravitational field equations are derived by varying $L_G$ with respect to $b^i$ and $\om^{ij}$.
The status of Riemannian solutions in the framework of PG was clarified by Obukhov \cite{x5}:
\bitem
\item[$\vartriangle$] Any solution of GR with a cosmological constant is a torsion-free solution of the general PG field equations.
\eitem
In particular, this is true for Kerr-AdS solutions, the subject of the present work.

In the Hamiltonian approach, the asymptotic charges are defined as certain boundary terms at spatial infinity, which make the associated canonical gauge generator $G$ differentiable \cite{x16}. By extending this construction, one can naturally introduce \emph{entropy as a boundary term} on horizon \cite{x10}. Consider a stationary, axisymmetric black hole, and let $\S$ be a spatial section of spacetime whose boundary consists of two components, one at infinity and the other at horizon, $\pd\S=S_\infty\cup S_H$. The asymptotic charges and  black hole entropy are defined by the variational equations for the respective boundary terms $\G_\infty$ and $\G_H$:
\bsubeq\lab{2.2}
\bea
&&\d\G_\infty=\oint_{S_\infty}\d B(\xi)\,,\qquad
       \d\G_H=\oint_{S_H} \d B(\xi)\,,                               \\
&&\d B(\xi):=\frac{1}{2}(\xi\inn\om^{ij})\d H_{ij}
             +\frac{1}{2}\d\om^{ij}(\xi\inn H_{ij})\, .
\eea
\esubeq
where $\xi$ is a Killing vector ($\pd_t$ or $\pd_\vphi$ on $S_\infty$, and a linear combination thereof on $S_H$, such that $\xi^2=0$),
and  $H_{ij}$ is the covariant momentum determined by the Lagrangian \eq{2.1},
\bea
H_{ij}&:=&\frac{\pd L_G}{\pd R^{ij}}=-2a_0\hd(b_ib_j)-2\aa_0(b_ib_j) \nn\\
 &&+2\sum_{n=1,4,6}\big[\hd\big(b_n\ir{n}R_{ij}\big)
                             +\bb_n\ir{n}R_{ij}\big]\, ,             \lab{2.3}
\eea
The operation $\d$ is assumed to be in accordance with the following two rules:
\bitem
\item[(r1)] The variation $\d\G_\infty$ is defined by varying parameters of the black hole state over the boundary $S_\infty$, leaving the background configuration fixed.\vsm
\item[(r2)] The variation $\d\G_H$ is defined by varying the parameters on horizon, but keeping surface gravity constant over the horizon, in accordance with the zeroth law.
\eitem
When there exist finite solutions for  $\G_\infty$ and $\G_H$ ($\d$-integrability), they are interpreted as thermodynamic charges. Their  values are strongly correlated to the adopted boundary conditions.

The boundary terms $\G_\infty$ and $\G_H$ in \eq{2.2} are introduced as apriori independent objects. However, if the canonical generator $G$ defining local symmeties of a black hole is differentiable, the corresponding boundary term $\G$ is not needed, it vanishes. Thus, assuming the boundary $S_H$ to have the opposite orientation with respect to $S_\infty$, we have
\be
\d\G:=\d\G_\infty-\d\G_H=0\,,                                        \lab{2.4}
\ee
which is nothing but the \emph{first law} of black hole thermodynamics.

As follows from Eq. \eq{2.3}, the covariant momentum $H_{ij}$ consists of two independent parts, defined by the even and odd parity sectors of $L_G$. Consequently, each of the boundary terms $\G_\infty$ and $\G_H$ can be represented as the sum of two parts with opposite parities.

\section{Kerr-AdS geometry}\label{sec3}
\setcounter{equation}{0}

\subsection{Tetrad formalism}

The Kerr-AdS metric is a solution of GR with a cosmological constant. In   Boyer-Lindquist coordinates $(t,r,\th,\vphi)$, it can be formulated in terms of the orthonormal tetrad \cite{x11,x12}
\bsubeq\lab{3.1}
\bea
&&b^0=N\left(dt+\frac{a}{\a}\sin^2\th\,d\vphi\right)\,,\qquad
  b^1=\frac{dr}{N}\,,                                                \nn\\
&&b^2=Pd\th\, ,\qquad
  b^3=\frac{\sin\th}{P}\left(a\,dt+\frac{(r^2+a^2)}{\a}d\vphi\right)\,,
\eea
where
\bea
&&N=\sqrt{\D/\r^2}\, ,\qquad \r^2=r^2+a^2\cos^2\th\,,                \nn\\
&&\D=(r^2+a^2)(1+\l r^2)-2mr\,,\qquad \a=1-\l a^2\,,                 \nn\\
&&P=\sqrt{\r^2/f}\,, \qquad  f=1-\l a^2\cos^2\th\, .
\eea
\esubeq
Here, $0\le\th<\pi$ and $0\le\vphi<2\pi$, $m$ and $a$ are parameters of the solution, and $\l$ is determiend by the PG field equations, $3a_0\l=-\L_0$.

The metric $ds^2=\eta_{ij}b^i\otimes b^j$, which is stationary and axially symmetric, admits the Killing vectors $\pd_t$ and $\pd_\vphi$. Many metric-related characteristics of geometry play an essential role in black hole thermodynamics, such as the location of the outher horizon  $r=r_+$, the horizon area $A_H$, the angular velocity $\om_+$ and the surface gravity $\k$,
\bsubeq
\bea
&&\D(r_+)\equiv(r_+^2+a^2)(1+\l r_+^2)-2mr_+=0\,,                \lab{3.2a}\\
&&A_H=\int_{r_+}b^2b^3=4\pi\frac{r_+^2+a^2}{\a}\,,                   \\
&&\om=\frac{g_{t\vphi}}{g_{\vphi\vphi}}\,,\qquad
  \om_+=\om|_{r_+}=\frac{a\a}{r_+^2+a^2}\,,          \\
&&\k=\frac{[\pd\D]_{r_+}}{2(r_+^2+a^2)}\,.
\eea
\esubeq
The quantities $\om_+$ and $\k$ are constant over the horizon, and for large $r$, $\om\sim-\l a$.

For a given tetrad field, one can calculate the Riemannian connection 1-form $\om^{ij}$ by
\be
\om^{ij}=\frac{1}{2}\Big[h^i\inn db^j-h^j\inn db^i
                        -\Big(h^i\inn(h^j\inn db^m)\Big)b_m\Big]\,.
\ee
Then, the curvature is defined by $R^{ij}=d\om^{ij}+\om^i{}_k\om^{kj}$, and it has only two nonvanishing irreducible parts, $\ir{1}R^{ij}$ and $\ir{6}R^{ij}$. The first part, also known as the Weyl curvature, $W^{ij}\equiv\ir{1}R^{ij}$, is given by \cite{x11}
\bsubeq\lab{3.4}
\bea
&&W_{01}=2Cb^0b^1+2D b^2b^3\,,\qquad W_{12}= Cb^1b^2-Db^0b^3 \,,    \nn\\
&&W_{02}=-Cb^0b^2+Db^1b^3\, ,\qquad W_{13}= Cb^1b^3 +D b^0b^2 \,,   \nn\\
&&W_{03}=-Cb^0b^3 -D b^1b^2\,,\qquad W_{23}=-2Cb^2b^3+2D b^0b^1\,,
\eea
where the coefficients $C$ and $D$ are the same as in Kerr spacetime \cite{x17},
\be
C:=\frac{mr}{\r^6}(r^2-3a^2\cos^2\th)\, ,\qquad
D:=\frac{ma\cos\th}{\r^6}(3r^2-a^2\cos^2\th)\, .
\ee
\esubeq
The second irreducible part is $\ir{6}R^{ij}=\l b^ib^j$.

For $m=0$, we have $W^{ij}\equiv R^{ij}-\ir{6}R^{ij}=0$, and the curvature corresponds to the background, AdS configuration, $R^{ij}=\l b^i b^j$.
The quadratic, even and odd curvature invariants,
\be
W^{ij}\,\hd W_{ij}=24 (C^2-D^2)\heps\,,\qquad W^{ij}W_{ij}=-48(CD)\heps\,,
\ee
are singular at $\r^2=0$ (a ring at $r=0,\th=\pi/2$).

\subsection{Limitations of Boyer-Lindquist coordinates}

The background configuration for $m=0$ is described by the AdS geometry but in somewhat \emph{non-standard coordinates}, in which metric components depend on the parameter $a$. Hence, one cannot clearly distinguish whether the variation $\d a$ is related to the background or to the genuine black hole configuration.
To avoid the variation of the AdS background, we introduce an improved version of the rule (r1).
\bitem
\item[(r1$^\prime$)] When $\d\G_\infty$ is calculated for Kerr-AdS black holes in Boyer-Lindquist coordinates, first apply $\d$ to all $a$'s, then remove those $\d a$ terms that survive the limit $m=0$, as they are associated to the background configuration.
\eitem

However, this is not sufficient to make the Boyer-Lindquist coordinates well defined. Namely, as shown by Henneaux and Teitelboim \cite{x17}, see also Carter \cite{x18}, the metric in these coordinates does not have a proper, asymptotically AdS behavior, needed for the canonical identification of asymptotic charges. To avoid the problem, they introduced a suitable change of coordinates which brings the metric to the standard, asymptotically AdS form. In our variational approach \eq{2.2}, the problem shows up as $\d$-nonintegrability of the asymptotic charges. It can be resolved by using the reduced form of the Carter-Henneaux-Teitelboim coordinate transformations, see for instance \cite{x11},
\be
T=t\, ,\qquad \phi=\vphi-\l a t\,.                                   \lab{3.6}
\ee
This change of coordinates and the improved rule (r1$^\prime$) ensure the new asymptotic charges, $\d E_T:=\d\G_\infty[\pd_T]$ and
$\d E_\phi:=\d\G_\infty[\pd_\phi]$, to be well defined. In addition to that,
\be
\d E_T=\d E_t+\l a\d E_\vphi\, ,\qquad \d E_\phi=\d E_\vphi\,.
\ee
Black hole entropy is determined by $\d\G_H[\xi]$, where $\xi=\pd_T-\Om_+\pd_\phi$ and
\be
\Om:=\frac{g_{T\phi}}{g_{\phi\phi}}=\om+\l a\,,\qquad \Om_+=\Om\big|_{r_+}\,.
\ee
For large $r$, the new angular velocity $\Om$ vanishes, as expected \cite{x9}. It turns out that the expression $\d\G_H[\xi]$ is invariant under the coordinate transformations \eq{3.6}.

In what follows, we will use the notation \pgp\ and \pgm\ for the even and odd parity sectors of PG, respectively.

\section{Thermodynamic charges in \pgp}\label{sec4}
\setcounter{equation}{0}

To simplify technical exposition of our analysis of Kerr-AdS thermodynamics, we first ana\-lyse Eqs. \eq{2.2} in the \pgp\ sector, leaving \pgm\ for the next section.

The \pgp\ sector is  effectively described by the Lagrangian
\bsubeq\lab{4.1}
\be
L^+_G=-\hd(a_0R+2\L)+\frac{1}{2}R^{ij}\hd (b_1 W_{ij}+b_6\ir{6}R_{ij})\,,
\ee
and the corresponding covariant momentum is
\bea
H_{ij}&=&-2a_0\hd(b_ib_j)+2\,\hd\big(b_1W_{ij}+b_6\ir{6}R_{ij}\big)  \nn\\
 &=&-2(a_0-\l b_6)\hd(b_ib_j)+2b_1\hd W_{ij}\,.                     \lab{4.1b}
\eea
\esubeq
The expressions for angular momentum, energy and entropy, produced by the first term in \eq{4.1b}, are of the GR form, but with $a_0\to(a_0-\l b_6)$, see  \cite{x11}:
\bsubeq\lab{4.2}
\bea
&&\d E^\gr_\phi=16\pi (a_0-\l b_6)\d\Big(\frac{ma}{\a^2}\Big)\,,   \lab{4.2a}\\
&&\d E^\gr_T=16\pi (a_0-\l b_6)\d\Big(\frac{m}{\a^2}\Big)\,,       \lab{4.2b}\\
&&\d\G^\gr_H=T\d S^\gr\,,\quad
        S^\gr=16\pi(a_0-\l b_6)\frac{A_H}{4}\,,                    \lab{4.2c}
\eea
\esubeq
where $T=\k/2\pi$. Hence, to obtain the complete result, one needs to calculate only an additional contributions from the Weyl curvature term $H^{W}_{ij}:=2b_1\hd W_{ij}$.

In the calculations that follow, the integration over the boundaries is implicit.

\subsection{Asymptotic charges}

\prg{Angular momentum.} We start with the additional contribution to angular momentum, determined by the $W$-reduced relation \eq{2.2} with $\xi=\pd_\phi=\pd_\vphi$,
\be
\d E^W_\vphi:=\d\G_\infty^W[\pd_\vphi]=\frac{1}{2}\om^{ij}{}_\vphi H^W_{ij}
  +\frac{1}{2}\d\om^{ij}H^W_{ij\vphi}\, .
\ee
Here, there are only two nonvanishing terms,
\bea
&&\om^{01}{}_\vphi\d H^W_{01}+\d\om^{01}H^W_{01\vphi}
  =4\l b_1\d\Big(\frac{ma}{\a^2}\Big)\sin^3\th d\th d\vphi\,,        \nn\\
&&\om^{13}{}_\vphi\d H^W_{13}+\d\om^{13}H^W_{13\vphi}
  =2\l b_1\d\Big(\frac{ma}{\a^2}\Big)\sin^3\th d\th d\vphi\,.        \nn
\eea
After completing the integration, one obtains
\be
\d E^W_\phi=\d E^W_\vphi=16\pi\l b_1\d\Big(\frac{ma}{\a^2}\Big)\,.
\ee
Summing up this expression with the GR-like term \eq{4.2a}, the complete \pgp\ contribution to angular momentum takes the form
\be
\d E_\phi=16\pi A_0\,\d\Big(\frac{ma}{\a^2}\Big)\,,\qquad
                A_0:=(a_0-\l b_6)+\l b_1\,.                          \lab{4.5}
\ee

\prg{Energy.} Consider now the $W$-contribution to energy, defined by the variational equation \eq{2.2} with $\xi=\pd_T=\pd_t+\l a\pd_\phi$,
\bsubeq
\bea
&&\d E_T^W:=\d\G_\infty^W[\pd_T]=\d E_t^W+\l a \d E_\vphi^W\, ,     \lab{4.6a}\\
&&\d E_t^W=\frac{1}{2}\om^{ij}{}_t\d H^W_{ij}
                                     +\frac{1}{2}\d\om^{ij}H^W_{ijt}\, .
\eea
\esubeq
There are three nonvanishing contributions to $\d E_t^W$, defined by $(i,j)=(0,1),(1,2)$ and $(1,3)$. A direct calculation yields the result
\be
\d E_t^W=16\pi\l b_1
 \Big[\d\Big(\frac{m}{\a}\Big) +\frac{m}{2}\d\Big(\frac{1}{\a}\Big)\Big]\,,
\ee
which is \emph{not $\d$-integrable}. Such an inconsistency of  Boyer-Lindquist coordinates was noted also in GR \cite{x11}. Transition to the well-behaved $(T,\phi)$ coordinates via \eq{4.6a} yields
\be
\d E_T^W=16\pi\l b_1\,\d\Big(\frac{m}{\a^2}\Big)\,.
\ee
Then, adding the GR-like term \eq{4.2b} yields the complete \pgp\ contribution to energy,
\be
\d E_T=16\pi A_0\,\d\Big(\frac{m}{\a^2}\Big)\,.                      \lab{4.9}
\ee

\subsection{Entropy and the first law}

For $\xi=\pd_T-\Om_+\pd_\phi$, the $W$-contribution to entropy is given by
\be
\d\G_H^W[\xi]:=\frac{1}{2}\om^{ij}{}_\xi \d H^W_{ij}
                      +\frac{1}{2}\d\om^{ij}H^W_{ij\xi}\, ,
\ee
where $\om^{ij}{}_\xi:=\xi\inn\om^{ij}$ and similarly for $H^W_{ij\xi}$. It contains only one nonvanishing contribution,
\bsubeq\lab{4.11}
\bea
\d\G_H^W[\xi]&=&
      \om^{01}{}_\xi \d H^W_{01}=4b_1\k\,\d\big(Cb^2b^3\big)     \lab{4.11a}\\
   &=&8\pi b_1\,\k\,\d\left(\frac{1+\l r_+^2}{\a}\right)
    =8\pi \l b_1\k\,\d\left(\frac{r_+^2+a^2}{\a}\right)\,.       \lab{4.11b}
\eea
\esubeq
In \eq{4.11a}, we used $\om^{01}{}_\xi=-\k$, and in \eq{4.11b}, the first term is obtained by integration, and the last equality follows from the identity
$1+\l r_+^2=\a+\l(r_+^2+a^2)$. Summing the above result with \eq{4.2c}, one obtains the complete \pgp\ expression for entropy,
\be
\d\G_H[\xi]=8\pi A_0\k\,\d\left(\frac{A_H}{4\pi}\right)=T\d S\,,\quad
    S:=16\pi A_0\,\frac{A_H}{4}\,.                                   \lab{4.13}
\ee

Each of the Kerr-AdS thermodynamic charges $(E_\phi,E_T,S)$ in the \pgp\ sector
can be obtained from the corresponding GR expression by $a_0\to A_0$. Hence, the first law is automatically satisfied,
\be
\d E_T-\Om_+\d E_\phi=T\d S\,.                                       \lab{4.14}
\ee

\section{Thermodynamic charges in \pgm}\label{sec5}
\setcounter{equation}{0}

The analysis starts with the effective Lagrangian
\bsubeq
\be
L^-_G=-\aa_0\hd X+\frac{1}{2}R^{ij}(\bb_1 W_{ij}+\bb_6\ir{6}R_{ij})\,,
\ee
and the corresponding covariant momentum
\bea
H_{ij}&=&-2\aa_0(b_ib_j)+2\,\big(\bb_1W_{ij}+\bb_6\ir{6}R_{ij}\big)\nn\\
 &=&-2(\aa_0-\l\bb_6)(b_ib_j)+2\bb_1 W_{ij}\,.                       \lab{5.1b}
\eea
\esubeq

\prg{1.} The \pgm\ expression for angular momentum is defined by
\bea
&&\d E_\vphi:=\d\G_\infty[\pd_\vphi]=\frac{1}{2}\om^{ij}{}_\vphi H_{ij}
  +\frac{1}{2}\d\om^{ij}H_{ij\vphi}\, .                              \lab{5.2}
\eea
Here, there are only two nonvanishing terms,
\bea
&&\om^{23}{}_\vphi\d H_{23}+\d\om^{23}H_{23\vphi}
 =\d\big[\om^{23}{}_\vphi( H_{23})_{\th\vphi}\big]d\th d\vphi\,,     \nn\\
&&\om^{02}{}_\vphi\d H_{02}+\d\om^{02}H_{02\vphi}\
  =\d\big[\om^{02}{}_\vphi (H_{02})_{\th\vphi} \big]d\th d\vphi\,.   \nn
\eea
For each of these terms, the integration over $\th$ yields an expression of the general form
\be
I=\int_0^\pi d\th f(\cos^2\th)\cos\th\sin\th\,,                    \lab{5.3}
\ee
where the change of variables $x=\cos\th$ implies $I=0$. Hence, the complete angular momentum stemming from the \pgm\ sector vanishes,
\be
\d E_\phi=\d E_\vphi=0\,.
\ee

\prg{2.} The \pgm\ contribution to energy is determined by the relations
\bsubeq
\bea
&&\d E_T=\d E_t+\l a\d E_\vphi=\d E_t\, ,                            \\
&&\d E_t:=\d\G_\infty[\pd_t]=\frac{1}{2}\om^{ij}{}_t \d H_{ij}
  +\frac{1}{2}\d\om^{ij}H_{ijt}\, .
\eea
\esubeq
There are seven nontrivial contributions to $\d E_t$,
\bea
&&\om^{01}{}_t \d H_{01}\,,\qquad \om^{23}{}_t \d H_{23}\,,
  \qquad \d\om^{23} H_{23t}\,,                                       \nn\\
&&\d\om^{0c} H_{0ct}\,,\qquad \d\om^{1c} H_{1ct}\,,\quad c=2,3\,.
\eea
A straightforward calculation shows that they are all of the general form \eq{5.3}, so that
\be
\d E_T=\d E_t=0\,.
\ee

\prg{3.} Consider the variational expression for $\d\G_H[\xi]$ with $\xi=\pd_T-\Om_+\pd_\phi$,
\be
\d\G_H[\xi]=\frac{1}{2}\om^{ij}{}_\xi \d H_{ij}
                      +\frac{1}{2}\d\om^{ij}H_{ij\xi}\, .
\ee
There is only one nontrivial term on the right-hand side,
\be
\d\G_H[\xi]=\om^{01}{}_\xi \d H_{01}=-4\bb_1\k\,\d\big(Db^2b^3\big)\,.
\ee
Again, the integral on the right-hand side is of the general type \eq{5.3}, hence, the \pgm\ contribution to black hole entropy vanishes,
\be
\d\G_H[\xi]=T\d S=0\,.
\ee

\section{Concluding remarks}\label{sec6}
\setcounter{equation}{0}

In the present paper, we analyzed thermodynamic properties of \emph{Riemannian} Kerr-AdS solutions in the context of general, parity-violating PG models, using the Hamiltonian approach proposed in \cite{x10}. Black hole entropy and asymptotic charges are completely determined by the contributions stemming from the \pgp\ sector, whereas those from the \pgm\ sector vanish. The general form of the thermodynamic charges guarantees that the first law is automatically satisfied.

Using the identity $W_{ij}=R_{ij}-\l b_ib_j$, one can rewrite the complete covariant momentum $\text{\eq{4.1b}}+\text{\eq{5.1b}}$ in an equivalent form as
\bea
H_{ij}&=&-2A_0\hd(b_ib_j)+2b_1\hd R_{ij}                               \nn\\
  &&-2\bA_0 (b_ib_j)+2\bb_1 R_{ij}\,.
\eea
where $\bA_0=(\aa_0-\l\bb_6)+\l\bb_1$.
The last terms in the upper and lower line are associated to the Euler and Pontryagin topological invariants, $R^{ij}\hd R_{ij}$ and $R^{ij}R_{ij}$, respectively, in the Lagrangian. Our calculations show that these two terms produce vanishing contributions to the termodynamic charges. Note also that the Holst term is not a topological invariant, but it also has no impact on the Kerr-AdS thermodynamic charges. These conclusions are in agreement with those obtained by Jacobson and Mohd \cite{x14} in their analysis of the tetrad form of higher curvature gravity.

Comparing the results obtained here to those describing  Kerr-AdS solutions with a \emph{nonvanishing torsion} \cite{x13}, one can conclude that they are characterized by different characteristic constants $A_0$ and $a_1$,  respectively,
\be
A_0\equiv a_0+\l (b_1-b_6)\,,\qquad a_1\equiv a_0-\l(b_4+b_6)\, .
\ee
This difference can be understood as a consequence of different dynamical settings in the two cases, or, more specifically, as an effect of torsion on the Riemann--Cartan connection.



\end{document}